\documentclass{article}

\usepackage{PRIMEarxiv}

\usepackage[utf8]{inputenc} 
\usepackage[T1]{fontenc}    
\usepackage{hyperref}       
\usepackage{url}            
\usepackage{booktabs}  
\usepackage{xcolor}
\usepackage{hyperref}
\usepackage{enumitem}
\usepackage{amssymb}
\usepackage{xcolor}
\usepackage{geometry}
\usepackage[round]{natbib}
\usepackage{xcolor}

\usepackage{etoolbox}
\definecolor{mybrown}{HTML}{662600}

\newcommand{\coloredcitep}[1]{(%
  \textcolor{mybrown}{\citeauthor{#1}, \citeyear{#1}}%
)}

\usepackage{amsfonts}       
\usepackage{nicefrac}       
\usepackage{microtype}      
\usepackage{lipsum}
\usepackage{fancyhdr}       
\usepackage{graphicx}       

\usepackage{tikz}
\usepackage{svg}
\usepackage{booktabs}
\usepackage{multirow}
\usepackage{graphicx}
\usepackage{subcaption}
\usepackage{siunitx}
\usepackage{colortbl}
\usetikzlibrary{automata, positioning, arrows.meta}

\usepackage{color}

\usepackage[table]{xcolor}
\usepackage[most]{tcolorbox}
\usepackage{pgfmath}

\definecolor{orange50}{HTML}{FFECE0} 
\definecolor{orange60}{HTML}{FFD6B3}
\definecolor{orange70}{HTML}{FFBB80}
\definecolor{orange80}{HTML}{FF9F4D}
\definecolor{orange90}{HTML}{FF841A}
\definecolor{orange100}{HTML}{CC4E00} 

\newtcbox{\asrbox}[2][]{enhanced, box align=base, rounded corners=southeast,
    colback=#1, colframe=#1, boxrule=0pt, arc=3pt, outer arc=3pt, top=0pt, bottom=0pt,
    left=0.5pt, right=0.5pt, boxsep=0.5pt, nobeforeafter, #2}

\newcommand{\asrcell}[1]{%
  \begingroup
  \edef\value{#1}%
  \ifdim\value pt<50pt
    #1%
  \else
    \ifdim\value pt<60pt
      \cellcolor{orange50}#1%
    \else
      \ifdim\value pt<70pt
        \cellcolor{orange60}#1%
      \else
        \ifdim\value pt<80pt
          \cellcolor{orange70}#1%
        \else
          \ifdim\value pt<90pt
            \cellcolor{orange80}#1%
          \else
            \cellcolor{orange90}#1%
          \fi
        \fi
      \fi
    \fi
  \fi
  \endgroup
}

\graphicspath{{media/}}     

\title{The Power of Backdoor Absorption in Community Training
}

\author{
Issam Seddik, Sami Souihi, Mohamed Tamaazousti, Sara Tucci Piergiovanni \\
Université Paris-Saclay, CEA LIST, Palaiseau, France \\
\texttt{\{issam.seddik, sami.souihi, mohamed.tamaazousti, sara.tucci\}@cea.fr}
}

\begin{document}
\maketitle

\begin{abstract}
Backdoor attacks severely threaten large-scale AI models. When model owners delegate training to external compute providers within a decentralized training paradigm, adversaries can craft stealthy, low-frequency triggers to inject malicious behavior while evading standard audits. Traditionally, detecting these attacks requires a full re-computation of the training steps—a prohibitive overhead that directly contradicts the owner's resource constraints.  To address this, we investigate the resilience of continuous optimization dynamics under Byzantine perturbations, where adversaries are forced to compete against a continuous influx of honest updates. Under a threat model where an adversary compromises $f$ out of $n$ total trainers, we quantify the minimum auditing overhead required by the model owner to probabilistically bound the attack success rate. We formalize this injection-absorption dynamic as a Discrete-Time Markov Chain (DTMC). Using this framework, we prove that the success probability of any bounded adversary asymptotically collapses to zero under a defense strategy combining natural absorption, a randomized scheduler, and lazy verification oracle. Empirical results demonstrate significant backdoor suppression with zero utility degradation even when invoking the verification oracle on merely 10\% of the total training steps. This approach yields a provably sound and computationally efficient defense for safety-critical AI.
\end{abstract}


\section{Introduction}

As AI systems continue to achieve unprecedented performance across increasingly complex tasks \coloredcitep{abramson2024accurate, trinh2024solving}, their integration into safety-critical applications has made security a paramount concern. Specifically, large-scale AI models are highly vulnerable to stealthy backdoor attacks, where adversaries covertly embed malicious behavior during the model's training \coloredcitep{gu2017badnets, chen2017targeted}.

Modern AI training unfolds over a sequence of discrete optimization steps. In a community training paradigm, a computationally constrained model owner delegates these steps to an external community of trainers \coloredcitep{murturi2023community, hegedHus2021decentralized, ormandi2013gossip}. Adversaries exploit this distributed setting by compromising a subset of trainers to inject malicious updates. To evade detection, attackers operate under strict parsimony—minimizing the trigger magnitude ($\delta$) and poisoning ratio ($\rho$) \coloredcitep{li2021invisible, zeng2023narcissus}. To be successful, the injected backdoor must also be sufficiently persistent to survive subsequent honest updates until the model is deployed \coloredcitep{hubinger2024sleeper, guo2025persistent}.

Existing defenses exhibit fundamental limitations when confronted with these parsimonious and persistent attacks. Sanitization, anomaly detection, and provenance-based methods fail because stealthy triggers either mimic benign semantic variance or require unrealistic access to private local data \coloredcitep{zhu2023gradient, baruch2019little, roy2022eiffel}. Consequently, exhaustive cryptographic auditing remains the only mathematically rigorous alternative. However, verifying every single training step of a large-scale model imposes a prohibitive computational overhead that directly contradicts the owner's resource constraints \coloredcitep{abbaszadeh2024zero, 11391059}. Adversaries exploit this asymmetry: by stretching their injection over multiple steps, they can effectively outlast any sparse or economically viable verification budget \coloredcitep{yao2019latent, souly2025poisoning}.

Recent research demonstrates that neural networks inherently exhibit \textit{Backdoor Absorption}; backdoors can be naturally "washed out" when subjected to a continuous influx of subsequent benign updates \coloredcitep{souly2025poisoning, kemker2018measuring, zhang2022neurotoxin}. In this study, we investigate how to harness this natural backdoor absorption to defend models. Our primary objective is to determine how a model owner can leverage this absorption phenomenon alongside minimal calls to a verification oracle to achieve definitive security without incurring prohibitive overhead.

\begin{figure}[t!]
    \centering
    \includegraphics[width=0.70\textwidth]{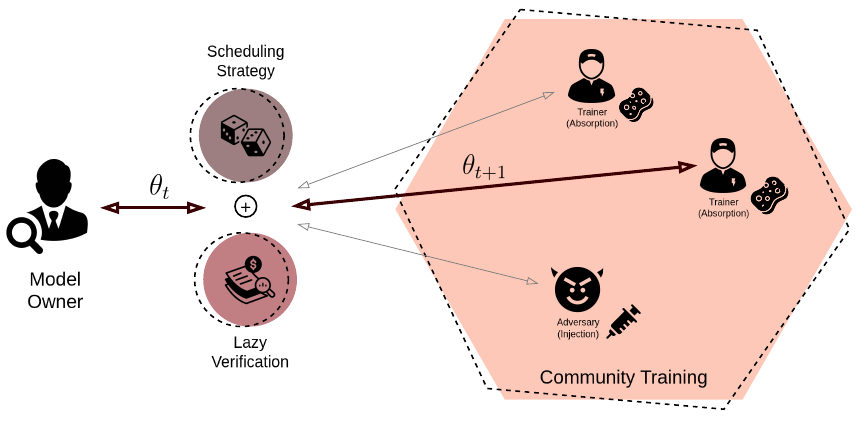}
    \caption{Adversarial injection and honest absorption dynamics in Community Training. The Model Owner delegates model parameters $\theta_t$ using randomized scheduling and lazy verification to participants. The solid arrow represents the chosen trainer selected to compute the subsequent model update $\theta_{t+1}$.}
    \label{fig:AL}
\end{figure}

To evaluate this, we model the training process within a community paradigm consisting of $n$ total trainers, where $f$ malicious trainers actively attempt backdoor injection. The model owner is equipped with three specific capabilities: reliance on the community's natural backdoor absorption, a dynamic scheduling strategy for trainer selection, and a lazy verification oracle (Figure \ref{fig:AL}). We rigorously formalize the stochastic tug-of-war between backdoor injection and benign absorption as a Discrete-Time Markov Chain (DTMC) \coloredcitep{ross2014introduction, levin2017markov}. Under these capabilities, we analytically quantify the minimum number of calls to the verification oracle the owner must make to dynamically penalize malicious trainers and drive the attack success probability asymptotically to zero ($\lim_{t \to \infty} p(t) = 0$).

We make the following contributions:

\noindent \textbf{C1)} We formally repurpose the backdoor absorption phenomenon as a defensive mechanism, proving that the natural accumulation of benign sequential updates induces a predictable degradation of parsimonious backdoors.

\noindent \textbf{C2)} We propose a closed-form DTMC framework that mathematically formalizes the interplay between backdoor injection and benign absorption. This model derives tight theoretical bounds on hitting times under both homogeneous and time-inhomogeneous regimes.

\noindent \textbf{C3)} We introduce a composite defense for community AI training that combines dynamic scheduling penalties with lazy verification. We demonstrate—both formally and empirically—that invoking the verification oracle on merely 10\% of the total training steps forces the attack success rate asymptotically to zero, neutralizing the threat with minimal computational overhead.

The rest of this paper is organized as follows: Section \ref{sec:background} provides background on AI training, backdoors, community training paradigms, and related work. Section \ref{sec:model} formalizes the System and Threat Models. Section \ref{sec:dtmc} develops the core Markov Chain formulation and our dynamic defense, alongside numerical analyses. Section \ref{sec:evaluation} presents our empirical evaluation. Finally, Sections \ref{sec:discussion} and \ref{sec:conclusion} discuss broader implications, future work, and conclude the paper.

\section{Background \& Related Work} \label{sec:background}

We begin with a brief background on AI training preliminaries, backdoor attacks, and the evolution of community training paradigms.

\subsection{AI Training Preliminaries}

In formal machine learning paradigms, a model parameterized by weights $\theta$ is optimized to map inputs $x$ to desired outputs $y$ \coloredcitep{Goodfellow-et-al-2016}. The available data is partitioned into a training set, $\mathcal{D}_{train}$, used to optimize the model parameters, and a testing set, $\mathcal{D}_{test}$, reserved to evaluate generalization on unseen data. The learning process seeks to minimize a predefined loss function $\mathcal{L}(\theta; x, y)$ over the training set. The model is optimized through a sequence of discrete training steps, utilizing an update rule such as Stochastic Gradient Descent (SGD), formally defined as \coloredcitep{bottou2010large}:

\begin{equation}
    \theta_{t+1}=\theta_t-\eta\nabla_\theta\mathcal{L}(\theta_t;\mathcal{B}_t)
\end{equation}

\noindent where $\eta$ denotes the learning rate, $\theta_t$ represents the model parameters at step $t$, and $\mathcal{B}_t\subset\mathcal{D}_{train}$ is the data batch sampled during that step. During each step, a trainer samples this batch to compute the loss gradient and updates the model weights. Training concludes after a designated number of steps $T$ or upon convergence, at which point accuracy and robustness are evaluated against $\mathcal{D}_{test}$.

\subsection{Backdoor Attacks}
Backdoor attacks that leverage training dynamics represent a severe security threat, allowing adversaries to covertly embed stealthy malicious behaviors into the model \coloredcitep{gu2017badnets, chen2017targeted, shafahi2018poison, li2021invisible, hubinger2024sleeper, souly2025poisoning}. The efficacy of a backdoor attack is governed by two intrinsic properties \coloredcitep{bagdasaryan2020backdoor, sun2019can}:
\begin{itemize}
    \item \textit{Trigger Magnitude $\delta$:}  The perceptible or structural strength of the backdoor trigger added to the input space, often bounded by an $L_p$ norm (e.g., $||\delta||_p \leq \epsilon$) to maintain input-level stealth.
    
    \item \textit{Poisoning Ratio $\rho$:} The proportion of compromised data points within the adversary's local batch.
\end{itemize}
\noindent Together, the tuple $(\delta, \rho)$ defines the aggressiveness of the malicious step. The adversary seeks to inject a backdoor such that the resulting model attains an Attack Success Rate (ASR) meeting or exceeding a predefined threshold $\alpha$ on triggered inputs, while preserving accuracy on clean inputs. Achieving full backdoor injection up to $ASR_{max}$ requires a minimum budget of $\tau$ consecutive malicious training steps \coloredcitep{zeng2023narcissus}.

However, backdoor persistence is inherently undermined by \textit{backdoor absorption} (also known as backdoor degradation) \coloredcitep{du2019lifelong, guo2025persistent, liu2024beyond, souly2025poisoning}. The continuous integration of benign updates during post-training phases induces a gradual, measurable degradation in backdoor efficacy. We characterize this degradation through a per-honest-step ASR decay factor, stochastically quantifying the proportional reduction in ASR following each clean update.

Building on this, we introduce the \textit{absorption lifespan}, denoted $\pi$, defined as the maximum number of consecutive honest training steps the global model can sustain before the ASR falls strictly below the target threshold $\alpha$ \coloredcitep{souly2025poisoning}. Once $\pi + 1$ consecutive honest training steps elapse, the implanted backdoor is eradicated and the adversary's progress is completely nullified. By formalizing this outcome, we reframe the intrinsic absorption dynamics of neural networks as an actively exploitable defensive mechanism against persistent threats.

\subsection{The Evolution to Community Training}
Backdoor attacks are especially severe in decentralized AI training. In Federated Learning \coloredcitep{mcmahan2017communication}, data remains on client devices while a central server aggregates updates. Gossip Learning extends this to a fully decentralized topology \coloredcitep{hegedHus2021decentralized}. Both paradigms share a critical limitation: the absence of verifiable data provenance \coloredcitep{jia2021proof, 11391059}. Because data never leaves local devices, the system cannot audit what informed any given update, precluding formal guarantees against data-level manipulation \coloredcitep{bagdasaryan2020backdoor}.

Recent frameworks address this through community-structured AI \coloredcitep{warnat2021swarm, murturi2023community, cai2025blockchain, zhao2020privacy}, enabling cooperative dataset verification against shared security standards \coloredcitep{shayan2020biscotti}. Yet, a fundamental limitation persists: collective auditability requires continuous, exhaustive verification to detect backdoors, which carries prohibitive computational costs \coloredcitep{shejwalkar2021manipulating}. Despite their emphasis on shared governance, community frameworks lack a mechanism to efficiently deploy these defenses against stealthy adversaries without incurring massive overhead \coloredcitep{fang2020local, tolpegin2020data}.

This defensive gap directly motivates our focus on a \textit{community training paradigm}. Rather than relying on continuous, exhaustive verification or mandating unilateral data access, a computationally constrained model owner treats the underlying verification method as an abstract, costly oracle. By strategically delegating sequential training steps to the external community and deploying this \textit{lazy verification} oracle as sparingly as possible, the owner imposes a strict operational penalty on the adversary. This budget constraint, combined with the absorption dynamics inherent to sequential learning, provides the theoretical foundation for the composite backdoor defense developed in subsequent sections.

\section{System \& Threat Models} \label{sec:model}


\subsection{System Model}
We assume a community training paradigm \coloredcitep{jia2021proof, liu2021zkcnn} where a computationally constrained model owner $\mathcal{O}$ delegates the training of a machine learning model $f(x, \theta) \equiv f_\theta$ to an external community of trainers. Rather than relying on a single third party, the owner utilizes a community of trainers to leverage the natural backdoor absorption phenomenon—where clean updates inherently dilute maliciously injected backdoors.

We formalize the system model as follows. Let the network consist of $n$ total trainers, categorized into $n - f$ honest trainers and $f$ malicious trainers. Honest participants dilute injected backdoors through clean training, while malicious participants attempt to inject backdoors into the global model.

Unlike conventional federated learning \coloredcitep{mcmahan2017communication, hegedHus2021decentralized}, where participants concurrently submit local updates over private datasets, we envisage a sequential training model orchestrated by the owner (Figure \ref{fig:AL}). The learning process unfolds over $T$ total steps, and our analysis focuses on a contiguous window of $k \leq T$ steps. At each step $t \in \{1, \ldots, T\}$, the model owner selects exactly one trainer to perform the model update.

\noindent \textbf{Model Owner Goal.} The model owner's primary objective is to successfully train a clean, high-utility global model while strictly minimizing the computational overhead associated with auditing external trainers. 

\noindent \textbf{Model Owner Strategy.} Rather than exhaustively verifying every submitted update, the owner aims to thwart backdoor injections through a cost-effective, composite defense strategy. This strategy relies on three pillars: the natural absorption of injected backdoors by honest updates, a random scheduling mechanism governing trainer selection, and a verification method to check the correctness of a given update. The precise scheduling strategy and lazy verification mechanism are detailed later in the defense specification (Section \ref{sec:advanced}).

Ultimately, we assess backdoor success rates against an owner operating under these specific constraints. Let $q$ represent the probability of selecting an honest trainer at step $t$, and $p = 1 - q$ the probability of selecting a malicious one.








\subsection{Threat Model} \label{sec:threat}

In our threat model, we consider an adversary who controls $f$ out of $n$ external trainer nodes within the community to inject a backdoor, yielding a malicious selection probability of $p$ at each training step. We characterize the attack's temporal dynamics through two key parameters: the duration of active interference and the subsequent persistence of the implanted backdoor \coloredcitep{wang2020attack}.

\noindent \textbf{Formalizing Temporal Lifespans.} The temporal execution and subsequent decay of the attack can be formalized using two distinct lifespans:

\begin{itemize}
    \item \textit{Injection Lifespan $\tau$:} This parameter represents the active phase of the attack. We define $\tau$ as the required budget of poisoned updates needed to successfully inject the backdoor. Importantly, the adversary has no agency over when they are selected to participate; they can only decide whether to inject a poisoned batch once the owner calls upon them. Therefore, accumulating an injection lifespan of $\tau$ consecutive steps is entirely contingent on the owner's stochastic scheduling strategy selecting malicious nodes sequentially from step $t_{start}$ to $t_{end}$. This parameter measures the adversary's operational cost and their cumulative exposure to the owner's lazy verification mechanism.
    
    \item \textit{Absorption Lifespan $\pi$:} This parameter measures the persistence of the vulnerability post-injection, often referred to as backdoor durability \coloredcitep{zhang2022neurotoxin}. Once the adversary halts their participation at $t_{end}$, the global model continues to accumulate benign updates from honest trainers, initiating a process of natural absorption where the backdoor signal degrades \coloredcitep{kemker2018measuring}. We define $\pi$ as the number of steps following $t_{end}$ during which the model's ASR remains above a critical efficacy threshold, $\alpha$. Formally, $\pi = t_{decay} - t_{end}$, where $t_{decay} = \min \{t > t_{end} \mid \text{ASR}(t) < \alpha\}$.
\end{itemize}

\noindent Note that the dynamics of the backdoor are heavily dictated by the non-linear correlations between the attack properties $(\delta, \rho)$ and the temporal lifespans $(\tau, \pi)$ \coloredcitep{wang2020attack, zhang2022neurotoxin, kemker2018measuring}.

\noindent \textbf{Adversary Goal.} The adversary's primary objective is to stealthily inject a backdoor into the global model such that, at the unknown deployment time, the model achieves an ASR no less than a target threshold $\alpha$ (i.e., ASR $\ge \alpha$) on trigger-bearing inputs, while preserving benign accuracy on clean inputs to evade the owner's auditing.

\noindent \textbf{Adversary Capabilities.} The adversary possesses the same system knowledge as honest trainers: full access to the model architecture, the training recipe, and the data batches. When selected by the owner's scheduling strategy, a malicious trainer can decide to substitute the data batch for a given step $t$ with a poisoned counterpart, injecting a trigger pattern $\delta$ at poisoning ratio $p$. However, because the adversary cannot dictate the selection schedule, achieving full backdoor injection up to the target ASR requires the adversary to be serendipitously selected by the owner for $\tau$ consecutive steps.

Crucially, we operate under the \textit{blind horizon assumption}. Because the model owner centrally orchestrates the training and keeps the scheduling strategy strictly private, the adversary naturally has no knowledge of the total number of training steps $T$ nor of the remaining steps $T-t_{current}$. Rather than being an artificial constraint, this inherent lack of visibility is a fundamental reality of the system and is critical to the efficacy of the owner's composite defense. If the adversary knew the exact deployment step $T$, they could bypass the natural absorption process entirely by executing a last-minute injection strategy (i.e., poisoning only the final steps right before deployment) \coloredcitep{yao2019latent, baruch2019little}. Forced by the blind horizon, the adversary must inject blindly and rely strictly on the backdoor's persistence parameter $\pi$ to survive subsequent honest updates until an unknown deployment time.

To formally characterize the adversary's success probability against the owner's composite defense, we introduce a \textit{Markov chain} formulation. The training trajectory of the global model is mapped onto a discrete state space, wherein transition probabilities are governed by the interplay of the adversary's malicious interventions, the community's natural absorption dynamics, the owner's scheduling strategy, and the lazy verification mechanism. This formulation yields a mathematically tractable, closed-form characterization of the adversary's capacity to successfully mount a backdoor attack within this community training paradigm.

\section{Markov Chain Formulation} \label{sec:dtmc}
To quantify the adversary's probability of success within this community training paradigm, we model the sequential updates as a Discrete-Time Markov Chain (DTMC) \coloredcitep{dieuleveut2020bridging, shirokoff2025convergence, even2023stochastic}. This framework yields formal upper and lower bounds (i.e., the best- and worst-case scenarios for the adversary), mapping the stochastic transitions between states of backdoor injection and absorption-induced immunity \coloredcitep{zhu2023gradient}.

\subsection{State Space Definition}

\noindent \textbf{Baseline Variables.} Recall that under the owner's scheduling strategy, the probability of selecting a malicious trainer at any step $t$ is $p$, and the probability of selecting an honest trainer is $q=1-p$. To establish rigorous bounds on the attack's progression, our DTMC formulation models $p$ and $q$ as stationary transition probabilities that remain constant across all state transitions. The adversary's objective is to drive the global model to an ASR that meets or exceeds the chosen threshold $\alpha$, which necessitates the successful execution of $\tau$ consecutive malicious training steps.

\noindent \textbf{State Space Formulation.} We define the state space of the DTMC to track the stochastic tug-of-war between adversarial injection and the community's benign absorption. The state space is logically partitioned into two distinct phases based on the ASR efficacy threshold $\alpha$:

\begin{itemize}

	\item \textit{Injection States ($I_0, \dots, I_{\tau-1}$):} This partition represents the phase of training where the backdoor is not yet functionally viable ($\text{ASR} < \alpha$). The subscript index $i$ in state $I_i$ tracks the accumulated malicious progress—specifically, the number of consecutive malicious steps the adversary has successfully chained together toward their required injection budget of $\tau$.

	\item \textit{Absorption States ($A_0, \dots, A_\pi$):} This partition represents the phase where the backdoor has been successfully injected and remains viable ($\text{ASR} \ge \alpha$). Here, the subscript index $j$ in state $A_j$ tracks the accumulated consecutive honest steps that are actively driving the model toward the natural absorption of the backdoor.

\end{itemize}

\noindent Mapping the training process onto this discrete Markov state space allows us to derive two complementary bounds: the expected hitting time for the adversary to successfully traverse the Injection States, and the expected hitting time for the honest network to traverse the Absorption States to achieve complete backdoor absorption.

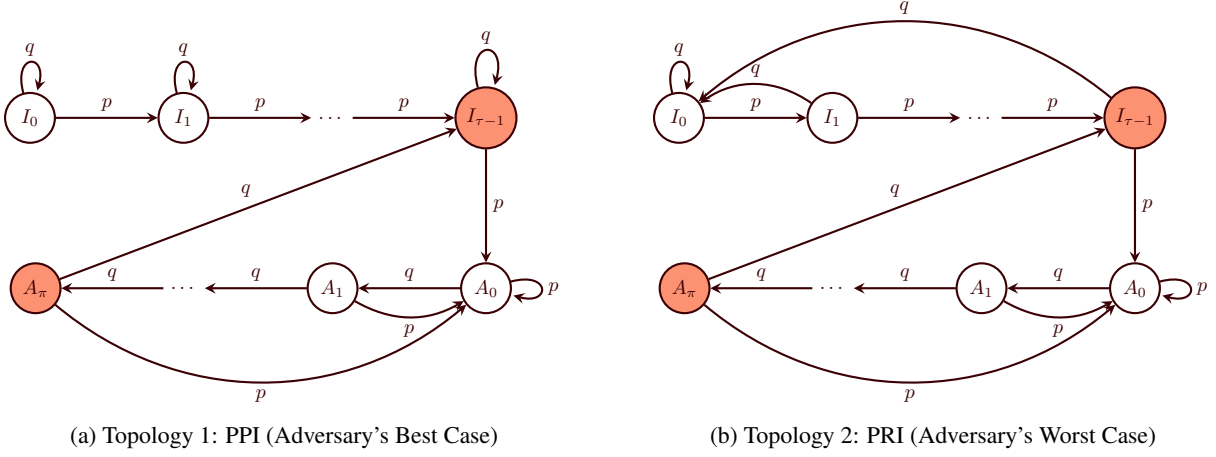
\begin{figure*}[t]
    \centering
    \definecolor{mahogany}{HTML}{3b0004}
    \definecolor{salmon}{HTML}{fc9272}
    
    \begin{subfigure}[b]{0.48\textwidth}
        \centering
        \begin{tikzpicture}[->, >=stealth, auto, thick, scale=0.75, every node/.style={transform shape}, color=mahogany, every state/.style={draw=mahogany, text=mahogany}]
            \node[state] (I0) {$I_0$};
            \node[state] (I1) [right=1.8cm of I0] {$I_1$};
            \node (dotsI) [right=1.8cm of I1] {$\dots$};
            \node[state, fill=salmon] (Itau) [right=1.8cm of dotsI] {$I_{\tau-1}$};

            \node[state] (A0) [below=2cm of Itau] {$A_0$};
            \node[state] (A1) [left=1.8cm of A0] {$A_1$};
            \node (dotsA) [left=1.8cm of A1] {$\dots$};
            \node[state, fill=salmon] (Api) [left=1.8cm of dotsA] {$A_\pi$};

            \path (I0) edge [loop above] node {$q$} (I0)
                  (I0) edge node {$p$} (I1)
                  (I1) edge [loop above] node {$q$} (I1)
                  (I1) edge node {$p$} (dotsI)
                  (dotsI) edge node {$p$} (Itau)
                  (Itau) edge [loop above] node {$q$} (Itau)
                  (Itau) edge node {$p$} (A0);

            \path (A0) edge [loop right] node {$p$} (A0)
                  (A0) edge node[above] {$q$} (A1)
                  (A1) edge [bend right=30] node[below] {$p$} (A0)
                  (A1) edge node[above] {$q$} (dotsA)
                  (dotsA) edge node[above] {$q$} (Api)
                  (Api) edge [bend right=40] node[below] {$p$} (A0)
                  (Api) edge node [above left] {$q$} (Itau);
        \end{tikzpicture}
        \caption{Topology 1: PPI (Adversary's Best Case)}
        \label{fig:topology1}
    \end{subfigure}
    \hfill
    \begin{subfigure}[b]{0.48\textwidth}
        \centering
        \begin{tikzpicture}[->, >=stealth, auto, thick, scale=0.75, every node/.style={transform shape}, color=mahogany, every state/.style={draw=mahogany, text=mahogany}]
            \node[state] (I0) {$I_0$};
            \node[state] (I1) [right=1.8cm of I0] {$I_1$};
            \node (dotsI) [right=1.8cm of I1] {$\dots$};
            \node[state, fill=salmon] (Itau) [right=1.8cm of dotsI] {$I_{\tau-1}$};

            \node[state] (A0) [below=2cm of Itau] {$A_0$};
            \node[state] (A1) [left=1.8cm of A0] {$A_1$};
            \node (dotsA) [left=1.8cm of A1] {$\dots$};
            \node[state, fill=salmon] (Api) [left=1.8cm of dotsA] {$A_\pi$};

            \path (I0) edge [loop above] node {$q$} (I0)
                  (I0) edge node {$p$} (I1)
                  (I1) edge [bend right=35] node[above] {$q$} (I0)
                  (I1) edge node {$p$} (dotsI)
                  (dotsI) edge node {$p$} (Itau)
                  (Itau) edge [bend right=40] node[above] {$q$} (I0)
                  (Itau) edge node {$p$} (A0);

            \path (A0) edge [loop right] node {$p$} (A0)
                  (A0) edge node[above] {$q$} (A1)
                  (A1) edge [bend right=30] node[below] {$p$} (A0)
                  (A1) edge node[above] {$q$} (dotsA)
                  (dotsA) edge node[above] {$q$} (Api)
                  (Api) edge [bend right=40] node[below] {$p$} (A0)
                  (Api) edge node [above left] {$q$} (Itau);
        \end{tikzpicture}
        \caption{Topology 2: PRI (Adversary's Worst Case)}
        \label{fig:topology2}
    \end{subfigure}
    
    \caption{DTMC state transitions for the adversary's injection phase, detailing (a) Progress-Preserving Injection (PPI) and (b) Progress-Resetting Injection (PRI). Both topologies share an absorption phase where malicious steps trigger complete progress resets.}
    \label{fig:topologies}
\end{figure*}

\subsection{Backdoor Injection Bounds}

We formally characterize the threat posed by the adversary during the injection phase by establishing tight mathematical bounds on the attack's progression. Because the global model is updated sequentially via the owner's scheduling strategy, the impact of an honest step occurring during the adversary's attempted injection traversal ($I_0 \rightarrow I_{\tau-1}$) defines the difficulty of the attack. We evaluate these bounds by analyzing the best-case and worst-case DTMC topologies from the adversary's perspective (Figure \ref{fig:topologies}).

\noindent \textbf{Topology 1: Progress-Preserving Injection (PPI)} \textit{(Adversary's Best Case).} In the most favorable scenario for the adversary, an honest step executed during the injection phase does not undo the adversary's partial progress (Figure \ref{fig:topology1}). In this topology, if the system is in state $I_i$ and an honest trainer is selected (probability $q$), the system simply undergoes a self-loop ($I_i \rightarrow I_i$), preserving the malicious gradient updates. The chain only advances ($I_i \rightarrow I_{i+1}$) when a malicious trainer is selected (probability $p$).

Because honest steps merely result in a self-loop, they do not reset the adversary's progress. Therefore, each of the $\tau$ required steps follows an independent geometric distribution with success probability $p$. By the linearity of expectation, the expected hitting time $\mu_I$—the total expected number of steps to inject the backdoor and cross from $I_0$ to the first absorption state $A_0$—is:

\begin{equation}
    \mu_I = \sum_{i=1}^{\tau} \frac{1}{p} = \frac{\tau}{p}
\end{equation}

\noindent \textbf{Topology 2: Progress-Resetting Injection (PRI)} \textit{(Adversary's Worst Case).} Conversely, in the strictest defensive scenario (Figure \ref{fig:topology2}), the natural absorption dynamics and the owner's verification force an immediate reset. A single honest step during the injection phase completely overwrites all preceding malicious progress. In this topology, any honest update (probability $q$) while in state $I_i$ immediately transitions the system back to the initial state $I_0$.To succeed under this topology, the adversary must achieve exactly $\tau$ consecutive malicious steps without a single honest interruption. If the adversary fails at step $k$ (where $1 \le k \le \tau$), which occurs with probability $p^{k-1}q$, progress is completely reset to $I_0$, and the remaining expected time becomes $\mu_I$ again. Solving the recursive expectation $\mu_I = \sum_{k=1}^{\tau} p^{k-1}q(k + \mu_I) + p^\tau(\tau)$ yields the bounded expected hitting time:

\begin{equation}
    \mu_I = \frac{1 - p^\tau}{q \cdot p^\tau}
\end{equation}

\subsection{Backdoor Absorption Bounds}

Once the adversary successfully executes the required $\tau$ malicious updates, the system breaches the target ASR threshold ($\text{ASR} \ge \alpha$) and transitions into the first absorption state, $A_0$. At this point, the backdoor is functionally viable. To permanently suppress this vulnerability and force the ASR below $\alpha$, the honest community must drive the natural absorption process, successfully traversing the entire sub-chain of absorption states from $A_0$ back down to the non-viable injection state $I_{\tau-1}$.

\noindent \textbf{Progress-Resetting Symmetry.} As formalized in Section \ref{sec:threat}, the absorption lifespan $\pi$ dictates the maximum number of honest steps the global model can endure before absorption degrades the trigger's efficacy. Consequently, to completely absorb the backdoor, the system must achieve exactly $\pi+1$ consecutive honest steps. Within our DTMC framework, every honest step (probability $q$) advances the absorption process ($A_j \rightarrow A_{j+1}$). However, the adversary does not remain passive; any malicious step (probability $p$) scheduled during this phase overwrites the natural decay, resetting progress entirely back to $A_0$. To ensure a worst-case security analysis, we maintain this absorption dynamic across all topologies—representing the attacker's optimal scenario—while focusing on bounding the injection phase. This dynamic exhibits a structural symmetry: the mathematical mechanics of the absorption phase are identical to the Progress-Resetting topology defined during the injection phase (Figure \ref{fig:topology2}). Here, the honest network attempts to maintain a consecutive streak of clean updates, while the adversary acts as the interrupter triggering immediate resets.

\noindent \textbf{Expected Absorption Time.} Leveraging this structural symmetry, we derive the expected absorption time bounds without redundant calculation by simply swapping the target sequence length ($\tau \leftrightarrow \pi+1$) and inverting the transition probabilities ($p \leftrightarrow q$) from the Progress-Resetting derivation. Applying this substitution yields the expected absorption time, $\mu_A$. This parameter represents the expected number of steps required for the community to traverse the absorption states and force the model back into the non-viable injection phase, despite the adversary's continuous attempts to interrupt the decay:

\begin{equation}
    \mu_A = \frac{1 - q^{\pi+1}}{p \cdot q^{\pi+1}}
\end{equation}

\noindent With both the expected injection time ($\mu_I$) and the expected absorption time ($\mu_A$) mathematically bound, we possess the complete set of parameters required to evaluate the overarching probability of adversarial success against the owner's composite defense over the training horizon.

\subsection{Asymptotic Vulnerability} \label{sec:proba}

Before introducing the owner's dynamic defense protocol, we first establish a theoretical baseline. Here, we derive the probability that the backdoor remains viable under a static scheduling strategy—where the selection probabilities ($p$ and $q$) remain constant and independent of the trainers' historical behavior. We analyze this over a known, finite number of steps before extending it to the infinite horizon setting necessitated by the blind horizon assumption.

\noindent \textbf{Finite Step Analysis.} We can find the exact probability of an active backdoor for a finite step $k$ using standard Markov Chain theory \coloredcitep{ross2014introduction, levin2017markov}. Let $P$ denote the transition matrix. This is a $(\tau+\pi+1)\times(\tau+\pi+1)$ matrix containing the $p$ and $q$ transition probabilities for all Injection and Absorption states defined in our topologies. Let $s_0$ be the initial state row vector representing the system's configuration at the start of training. Following the Chapman-Kolmogorov equations, after exactly $k$ steps, the probability distribution across all states is given by:

\begin{equation}
    s_k = s_0 \cdot P^k
\end{equation}

\noindent The adversary succeeds if the system is in any Absorption state ($A_0, \dots, A_\pi$) at step $k$. Therefore, the probability of a successful attack at a specific finite step is the sum of the probabilities of residing in the viable partition:

\begin{equation}
    P(\text{Success} \mid k) = \sum_{j=0}^{\pi} (s_k)_{A_j}
\end{equation}

\noindent \textbf{Asymptotic Success.} Under the practical constraints of the blind horizon assumption, the final training step $T$ is unknown to the adversary. Consequently, evaluating the long-term, asymptotic behavior of the global model's vulnerability is paramount. To evaluate this steady-state behavior, we must identify the correct renewal cycle of the Markov chain.

Crucially, when the system successfully absorbs the backdoor (exiting state $A_\pi$), the accumulated malicious progress is not entirely erased; rather, the system drops to the precipice of injection, state $I_{\tau-1}$. Let $\mu_I^\prime$ denote the expected renewal time—the expected number of steps required to return from the post-absorption state $I_{\tau-1}$ back into the functionally viable phase $A_0$. This renewal time depends entirely on the defensive topology:

\begin{itemize}
\item \textit{For PPI:} Because honest steps at $I_{\tau-1}$ simply result in a self-loop, this state functions as the exact regeneration point for our i.i.d. cycles within the PPI topology. The adversary requires only a single successful malicious step to re-inject the backdoor. Thus, $\mu_I^\prime = \frac{1}{p}$.
\item \textit{For PRI:} A successful step ($p$) immediately re-injects the backdoor, while an honest step ($q$) forces a complete reset back to $I_0$, requiring the adversary to wait the full expected hitting time ($\mu_I$) to climb back. This expected renewal time gracefully simplifies to $\mu_I^\prime = \frac{1}{p^\tau}$.
\end{itemize}

\noindent To evaluate the long-term threat, we define a reward process $R(t) = 1$ if the chain is in any Absorption state at step $t$, and $0$ otherwise. By applying the Renewal Reward Theorem \coloredcitep{gallager2013stochastic}, the long-run probability of being in the viable phase equals the expected viable time per cycle ($\mu_A$) divided by the total expected cycle length ($\mu_A + \mu_I^\prime$). We express the asymptotic success probability as:

\begin{equation}
   P_{\text{steady}} = \lim_{k \to \infty} P(\text{Success}) = \frac{\mu_A}{\mu_A + \mu_I^\prime}
\end{equation}

This baseline analysis exposes a critical vulnerability: in a static environment where the scheduling probabilities $p$ and $q$ remain constant, an adversary operating over an infinite time horizon will inevitably succeed for a predictable fraction of the training lifecycle. To definitively suppress this asymptotic vulnerability, the model owner must break the static nature of the transition matrix. In the subsequent sections, we detail the owner's dynamic defense protocol—incorporating lazy verification and adaptive scheduling penalties.

\subsection{Time-Inhomogeneous Defense} \label{sec:advanced}

While the baseline static analysis demonstrates that an adversary can theoretically succeed over an infinite horizon (Section \ref{sec:proba}), the owner's composite defense neutralizes this asymptotic threat by rendering the global model's transition matrix dynamic through adaptive scheduling and lazy verification.

\noindent\textbf{Lazy Verification.} Exhaustive verification of every gradient update submitted by the community is computationally prohibitive for the model owner. Accordingly, the owner adopts a lazy verification mechanism—a form of selective auditing \coloredcitep{11391059}—in which the owner audits training steps with probability $v$ and successfully detects malicious deviations with a detection rate $d$. This constraint structurally alters the transition probabilities. The effective probability of a successful malicious step drops to $p' = p(1 - v \cdot d)$, while the probability of an honest outcome (either a naturally clean update or a successfully caught and rejected malicious update) increases to $q' = q + p(v \cdot d)$.

\noindent\textbf{Dynamic Scheduling \& Penalties.} To systematically penalize malicious trainers identified by the lazy verification mechanism, the owner employs a dynamic, weight-based scheduling strategy. At any step $t$, the probability of designating a specific trainer is dynamically tied to its maintained trust weight. If the adversarial coalition controls an aggregate weight $W_{\mathrm{mal}}(t)$ and the honest trainers control an aggregate weight $W_{\mathrm{hon}}(t)$, the adversary's probability of being scheduled at step $t$ is:

\begin{equation}
    p_{\mathrm{sched}}(t) = \frac{W_{\mathrm{mal}}(t)}{W_{\mathrm{mal}}(t)+W_{\mathrm{hon}}(t)}
\end{equation}

Whenever lazy verification detects a malicious update, the owner immediately penalizes the offending trainer by reducing its weight. Using a predefined penalty fraction $S \in [0,1]$, the caught trainer's weight is updated as:

\begin{equation}
W_i(t+1) = (1-S) \cdot W_i(t)
\end{equation}

This penalty monotonically reduces the adversary's future selection probability.

\noindent\textbf{Time-Inhomogeneous Transition Matrix.} Because the penalty mechanism strictly decreases $W_{\mathrm{mal}}(t)$ upon each successful detection, the adversary's selection probability $p(t)$ becomes non-stationary. Consequently, the training process must be modeled as a \textit{time-inhomogeneous Markov chain} \coloredcitep{seneta2006non}, where the transition matrix $P^{(t)}$ evolves at every step. The state distribution after $k$ steps is no longer derived via simple matrix exponentiation, but rather by the ordered product of these evolving matrices:

\begin{equation}
\mathbf{s}_k = \mathbf{s}_0 \cdot \prod_{t=1}^{k} P^{(t)}
\end{equation}

\noindent\textbf{Asymptotic Collapse.} This inhomogeneous formulation provides the core security guarantee of this community training paradigm. As the owner's lazy verification triggers progressive penalties, the adversary's aggregate weight is structurally depleted, causing their scheduling probability to collapse asymptotically:

\begin{equation}
\lim_{t \to \infty} W_{\mathrm{mal}}(t) = 0 \;\implies\; \lim_{t \to \infty} p(t) = 0
\label{eq:collapse}
\end{equation}

Unlike static Markovian models—which remain vulnerable to persistent adversaries over an infinite horizon—the time-inhomogeneous dynamics guarantee that the attack success probability ultimately converges to zero. Critically, because this defense strategy relies on the natural absorption of the community paired with sparse auditing rather than universal inspection, the owner achieves definitive security without sacrificing training efficiency.

\subsection{Numerical Analysis} \label{sec:analysis}
We numerically evaluate the backdoor absorption dynamics within  the community training paradigm by simulating the DTMC state  transitions defined in Section~\ref{sec:dtmc}, quantifying  attack success probabilities across varying adversarial  capabilities, injection topologies, and defense configurations.

\begin{figure*}[t!]
    \centering
    \includegraphics[width=0.85\textwidth]{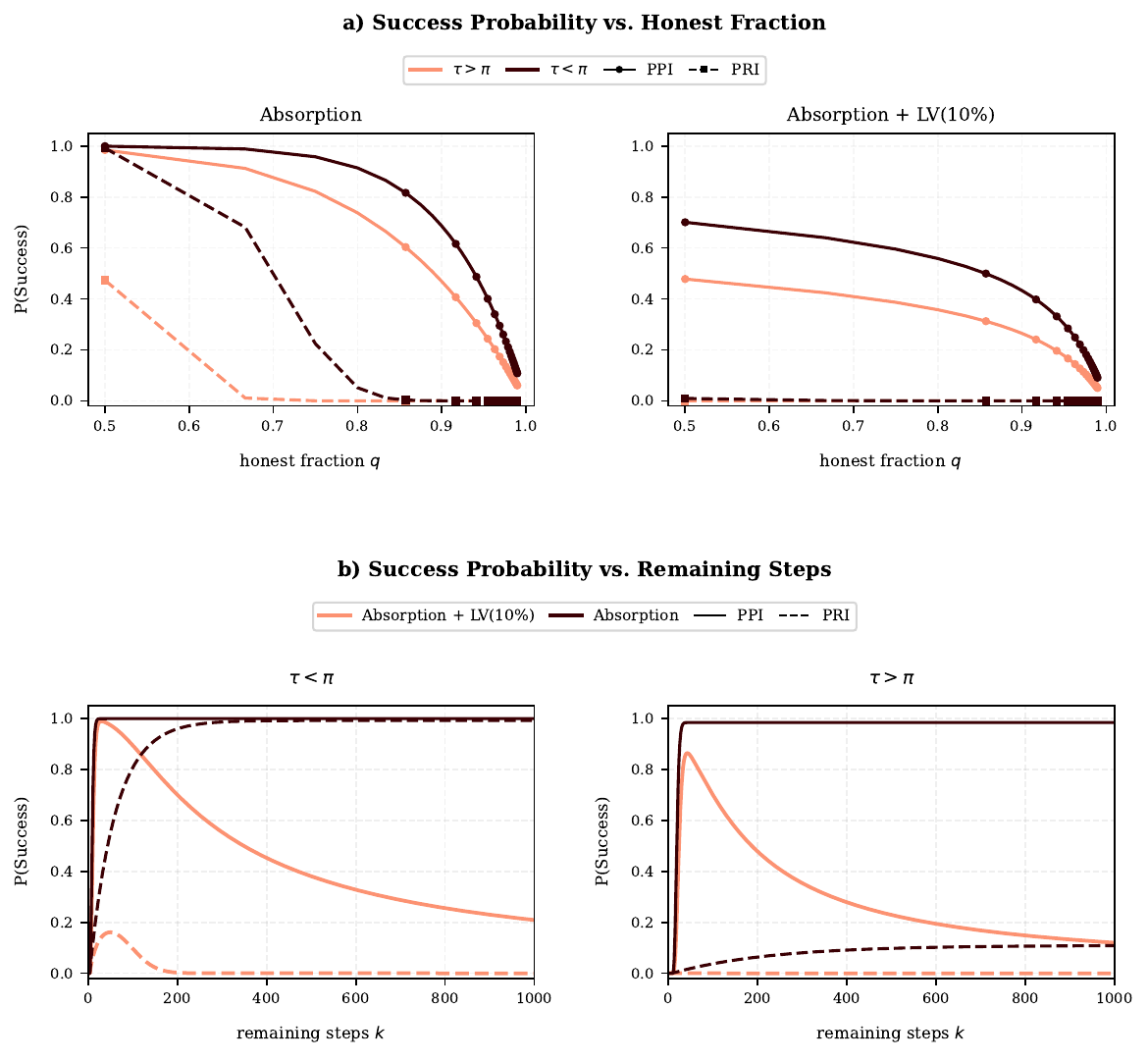}
    \caption{Theoretical backdoor absorption dynamics. (a) Attack success probability versus honest fraction $q$ $(k=200)$, comparing pure absorption to 10\% lazy verification. (b) Success probability over remaining steps $k$ $(p=q=0.5)$ across adversarial injection ($\tau$) and absorption ($\pi$) regimes under PPI and PRI topologies.}
    \label{fig:theoryres}
\end{figure*}

\begin{table}[t]
  \centering
  \caption{Analytical expected hitting times $(\mu_{I}^{\prime},\mu_{A})$ and steady-state success probabilities $(P_{steady})$ under baseline absorption across varying adversarial geometries $(\tau, \pi)$ and population dominances $(p, q)$.}
  \label{tab:detailed_hitting_times}
  
  \small
  
  \setlength{\tabcolsep}{2pt}
  
  \renewcommand{\arraystretch}{0.85}
  
  
  \begin{tabular}{@{}llcrrrr@{}}
    \toprule
    \textbf{Setting} $q$
      & \textbf{Regime}
      & \textbf{Topology}
      & $\mu_I^\prime$
      & $\mu_A$
      & $P_{\text{steady}}$ \\
    \midrule
    \multirow{4}{*}{\shortstack[l]{Absolute Parity\\($q=0.5,\;p=0.5$)}}
      & \multirow{2}{*}{$\tau > \pi$} & PPI & 20      & 126     & 98.4\,\% \\
      &                               & PRI & 2{,}046 & 126     & 11.0\,\% \\
    \cmidrule(lr){2-6}
      & \multirow{2}{*}{$\tau < \pi$} & PPI & 10      & 4{,}094 & 99.9\,\% \\
      &                               & PRI & 62      & 4{,}094 & 99.2\,\% \\
    \midrule
    \multirow{4}{*}{\shortstack[l]{Honest Dominant\\($q=0.90,\;p=0.10$)}}
      & \multirow{2}{*}{$\tau > \pi$} & PPI & 100    & 8.8 & 46.9\,\% \\
      &                               & PRI & ${\approx}1.11\times 10^{10}$ & 8.8 & $0.0$\,\% \\
    \cmidrule(lr){2-6}
      & \multirow{2}{*}{$\tau < \pi$} & PPI & 50      & 21.9  & 68.6\,\% \\
      &                               & PRI & 111{,}110 & 21.9  & 0.02\,\%  \\
    \midrule
    \multirow{4}{*}{\shortstack[l]{Malicious Dominant\\($q=0.10,\;p=0.90$)}}
      & \multirow{2}{*}{$\tau > \pi$} & PPI & 11.1 & 1{,}111{,}110 & 100.0\,\% \\
      &                               & PRI & 18.7   & 1{,}111{,}110 & 100.0\,\% \\
    \cmidrule(lr){2-6}
      & \multirow{2}{*}{$\tau < \pi$} & PPI & 5.6  & ${\approx}1.11\times 10^{11}$ & 100.0\,\% \\
      &                               & PRI & 6.9 & ${\approx}1.11\times 10^{11}$ & 100.0\,\% \\
    \bottomrule
  \end{tabular}
\end{table}

\noindent\textbf{Simulation Setup.} The system state tracks residual backdoor strength, bounded by the injection budget  $\tau$ and absorption lifespan $\pi$. We evaluate two capability  regimes ($\tau > \pi$ and $\tau < \pi$) under both injection  topologies (PPI and PRI), over a training horizon of  $k \leq 1000$ steps.

\noindent\textbf{Static Baseline \& Hitting Times.} Under pure  community absorption (i.e., without owner verification) selection  probabilities are stationary: $p = f/n$ and $q = (n-f)/n$.  Table~\ref{tab:detailed_hitting_times} reports the analytically derived  expected hitting times ($\mu_I'$, $\mu_A$) and steady-state  success rates ($P_{\text{steady}}$) across three population  regimes. The results expose extreme sensitivity to adversarial  representation. In the honest-dominant setting  ($q=0.90$, $p=0.10$), $\mu_I'$ surges while $\mu_A$ collapses; under PRI with $\tau > \pi$, $P_{\text{steady}}$ falls to 
$0.0\%$ (and $0.02\%$ for $\tau < \pi$). In the malicious-dominant setting ($q=0.10$, $p=0.90$), the adversary fully overwhelms absorption mechanics, achieving  $P_{\text{steady}} = 100\%$ across all topologies and regimes.

\noindent\textbf{Injection ($\tau$) vs.\ Absorption ($\pi$).} The  $\tau$-vs-$\pi$ relationship governs a structurally asymmetric competition between backdoor injection and honest absorption,  as visualized in Figure~\ref{fig:theoryres}b and  corroborated by Table~\ref{tab:detailed_hitting_times}. Under absolute  parity ($p = q = 0.5$), the two regimes produce sharply  divergent outcomes. When $\tau < \pi$ (e.g., $\tau=5$,  $\pi=10$), rapid injection paired with slow absorption sustains the backdoor almost indefinitely, driving $P_{\text{steady}}$  to $99.9\%$ under PPI and $99.2\%$ under PRI at $k=1000$ steps.  Forcing the adversary into $\tau > \pi$ (e.g., $\tau=10$,  $\pi=5$) renders PRI particularly hostile: the requirement for long uninterrupted malicious sequences reduces  $P(\text{Success})$ to $11.0\%$. This suggests that the optimal strategy for the adversary is to minimize $\tau$ relative to $\pi$.

\noindent\textbf{Asymptotic Collapse Under Active Defense.}  Figure~\ref{fig:theoryres}a contrasts $P(\text{Success})$ under pure absorption against the composite defense augmented  with Lazy Verification at a $10\%$ auditing budget  ($LV(10\%)$). Whereas pure absorption leaves the model broadly vulnerable across the honest fraction spectrum, $LV(10\%)$  structurally suppresses attack success across all values of $q$.  Figure~\ref{fig:theoryres}b further captures the temporal dynamics: under the static baseline at parity ($p=q=0.5$), the $\tau < \pi$ regime sustains the backdoor persistently over the full $k$-step horizon. Activating $LV(10\%)$ breaks this persistence, inducing rapid and irreversible collapse of  $P(\text{Success}) \to 0$ regardless of injection  topology---directly validating the asymptotic guarantee  established in Equation~\ref{eq:collapse}.

\begin{figure*}[ht!]
    \centering
    \includegraphics[width=0.99\textwidth]{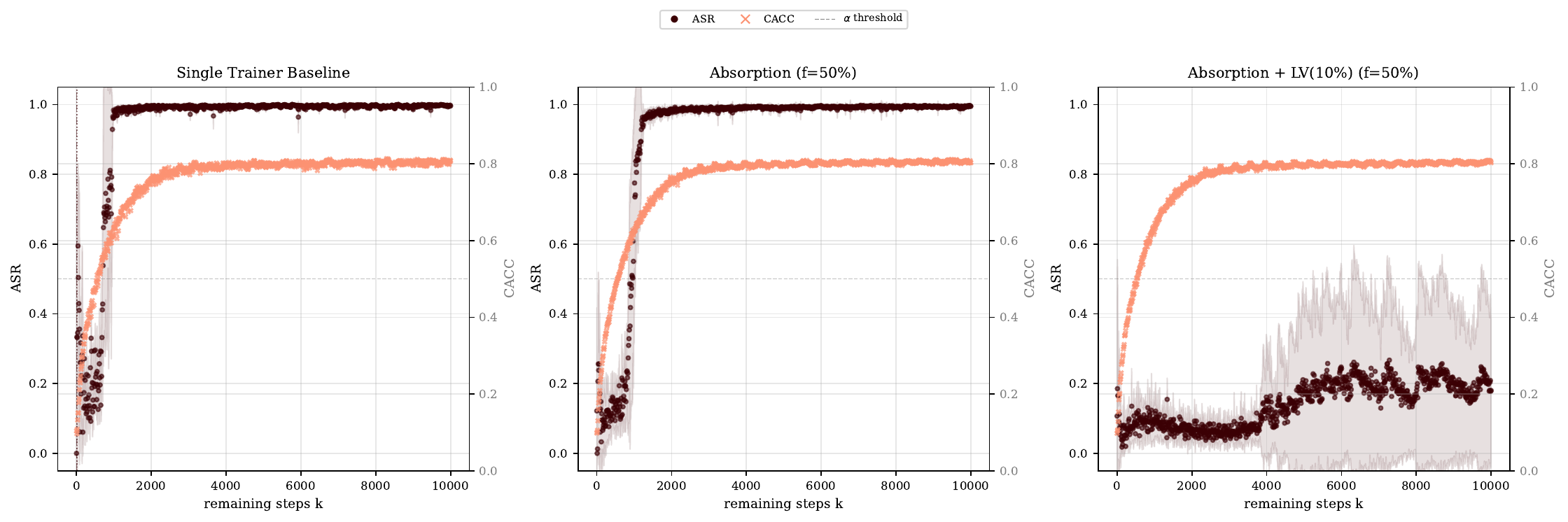}
    \caption{Attack Success Rate (ASR) and Clean Accuracy (CACC) evolution for a BadNet attack on ResNet-18 over 10,000 remaining steps. Panels contrast a single trainer baseline with community absorption $(f=50\%)$ and a composite defense combining absorption with 10\% lazy verification.}
    \label{fig:bcollapse}
\end{figure*}

\section{Empirical Evaluation} \label{sec:evaluation}

\subsection{Experimental Setup}

\noindent\textbf{Model \& Dataset.} We evaluate the backdoor absorption dynamics using a ResNet-18 architecture \coloredcitep{he2016deep} trained on the CIFAR-10 dataset \coloredcitep{krizhevsky2009learning}. This setup allows us to empirically characterize the stochastic tug-of-war between adversarial injection and benign absorption described in Section \ref{sec:model}.

\noindent \textbf{Attack Configuration.} We implement a BadNet attack \coloredcitep{gu2017badnets} to embed malicious behavior into the global model. Using a $3 \times 3$ pixel patch with a poisoning ratio of $p=0.3$ per malicious batch, we transition the model from source class ($5$) to target class ($2$) in CIFAR-10 dataset. To trace the theoretical bounds established in Section \ref{sec:dtmc}, we calibrate the injection budget ($\tau$) and absorption lifespan ($\pi$) across multiple experimental runs. The Attack Success Rate (ASR) viability threshold is set at $\alpha=0.5$.

\noindent \textbf{Community Training Environment.} The simulated community training environment consists of $N=100$ participant nodes. We evaluate the system under varying levels of adversarial dominance by manipulating the fraction of honest trainers ($q$) and malicious trainers ($f$), including a highly compromised regime where $f=50\%$. Training horizons are evaluated across sequences of $K=1,000$ and $K=10,000$ steps. To account for stochastic variance, all empirical metrics are averaged over 10 independent simulation runs. These configurations allow for the empirical confirmation of the asymptotic success probability bounds.

\noindent \textbf{Defense Configurations.} We isolate the effects of natural backdoor decay and our proposed verification strategy across three training configurations:(a) \textit{Baseline (Single Trainer):} Tracks standard attack progression without community absorption dynamics or verification.(b) \textit{Absorption Only:} Simulates a highly compromised network ($f=50\%$) where backdoor decay relies strictly on the continuous updates of honest trainers, without auditing interventions.(c) \textit{Absorption + Lazy Verification (LV):} Evaluates our composite defense under identical compromise conditions ($f=50\%$). This setup augments natural absorption with a Lazy Verification oracle that audits 10\% of training steps ($v=0.1$); detected malicious updates incur a weight penalty under the $S=0.2$ configuration. While our defense framework is fundamentally oracle-agnostic, our evaluation models the verification oracle after PoTS \coloredcitep{11391059}, which reduces per-step verification time to $30\%$ while maintaining a $0.9$ detection rate. Accordingly, we configure our simulated oracle with $d=0.9$. Because the internal mechanics of the oracle are orthogonal to our system-level contribution, our framework remains broadly applicable and will natively benefit from future verification protocols striving for deterministic accuracy ($d=1$).

\noindent \textbf{Evaluation Metrics.} Throughout the training trajectory, we continuously tracked two primary metrics: Clean Accuracy (CACC) to measure the model's benign utility, and Attack Success Rate (ASR) to quantify the viability of the injected backdoor against the $\alpha$ threshold.

\begin{figure}[ht!]
    \centering
    \includegraphics[width=0.45\textwidth]{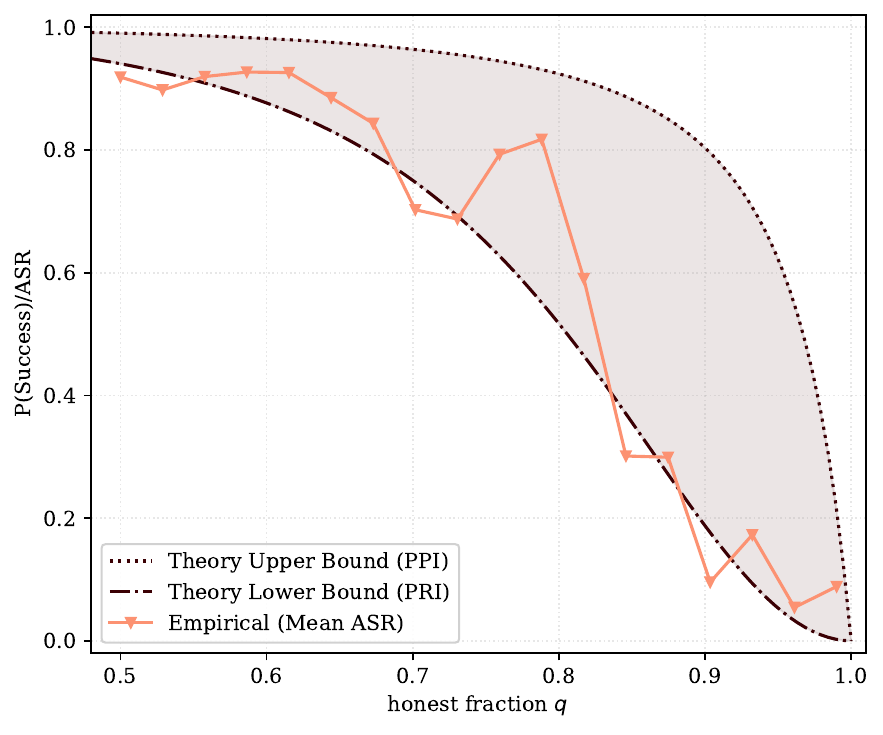}
    \caption{Empirical ASR evolution versus theoretical bounds for a BadNet attack on ResNet-18 over 1,000 remaining steps under absorption and lazy verification $(v=0.1, d=0.9, S=0.2)$. Empirical means remain bounded between the theoretical PPI upper limit and PRI lower limit, calibrated at $\tau=2$ and $\pi=27$.}
    \label{fig:theopra}
\end{figure}

\subsection{Results}

\begin{table*}[htbp]
    \centering
    \caption{Comparative performance and computational overhead, detailing Attack Success Rate (ASR), Clean Accuracy (CACC), and execution times across baseline, absorption-only $(f=0.5)$, and absorption with 10\% lazy verification regimes.}
    \label{tab:performance_comparison}
    \setlength{\tabcolsep}{6pt}
    \renewcommand{\arraystretch}{1.2}
    \begin{tabular}{
        l
        S[table-format=3.2]
        S[table-format=2.2]
        S[table-format=3.2]
        S[table-format=1.2]
        S[table-format=3.2]
    }
        \toprule
        \textbf{Scenario}
            & {\textbf{ASR} (\%)}
            & {\textbf{CACC} (\%)}
            & {\textbf{Base} (s)}
            & {\textbf{Overhead} (s)}
            & {\textbf{Total} (s)} \\
        \midrule
        Single Trainer Baseline
            & 99.80 & 79.65 & 184.89 & {---} & 184.89 \\
        Absorption Only ($f{=}50\%$)
            & 99.52 & 80.09 & 184.71 & {---} & 184.71 \\
        Absorption + LV(10\%) ($f{=}50\%$)
            & \textbf{7.20}  & \textbf{80.06} & 187.37 & \textbf{5.17} & 192.54 \\
        \bottomrule
    \end{tabular}
\end{table*}

\noindent \textbf{Backdoor Collapse.} Figure \ref{fig:bcollapse} presents the ASR and CACC evolution for a BadNet attack on ResNet-18 over $k=10{,}000$ steps. In the Single Trainer Baseline, the ASR persistently oscillates above the $\alpha=0.5$ threshold, showing the adversary maintains sustained injection capability without community dynamics. Under Absorption Only ($f=50\%$), continuous honest updates cause intermittent ASR suppression; however, the backdoor repeatedly resurges, proving that natural absorption alone cannot secure a highly compromised network. Conversely, Absorption + LV(10\%) achieves significant suppression: the ASR rapidly collapses below the viability threshold and remains stable, confirming the asymptotic guarantee of Equation \ref{eq:collapse}. Across all configurations, CACC remains high and unaffected, demonstrating that our composite defense neutralizes backdoors with zero utility degradation despite extreme adversarial compromise ($f=50\%$) and a minimal 10\% verification budget.

\noindent\textbf{Upper \& Lower Bounds Confirmation.} Figure \ref{fig:theopra} validates our Markov chain formulation: the empirical Mean ASR for a BadNet attack on ResNet-18 over $k=1{,}000$ steps remains bounded between the theoretical PPI upper and PRI lower limits. This gap exists because our analytical models isolate absolute best- and worst-case conditions, whereas actual stochastic training dynamics simultaneously delay adversarial injection and accelerate backdoor absorption. As the honest fraction ($q$) increases, the empirical ASR declines sharply toward the PRI lower bound, pushing real-world dynamics toward the adversary's worst-case scenario (the marginal non-zero ASR observed at $q=1.0$ stems from inherent training stochasticity). Conversely, highly compromised networks ($f=50\%$) track the PPI upper bound, sustaining near-maximal attack success. Ultimately, even a modest increase in honest participation triggers a steep, nonlinear drop in ASR, highlighting the disproportionate defensive value of recruiting honest trainers into the community.

\noindent\textbf{Overhead Analysis.} Table \ref{tab:performance_comparison} details the comparative performance and computational overhead across the three training regimes. CACC remains stable across all configurations (79.65\% to 80.06\%), confirming our defense strategy preserves benign model utility. Without verification, Absorption-Only setups ($f=50\%$) fail to suppress the threat, yielding near-total compromise (ASRs of 99.80\% and 99.52\%) with execution times around 184.8 s. In contrast, the composite defense (Absorption + 10\% LV) reduces the ASR to 7.20\%, achieving near-complete backdoor suppression. While traditional setups require exhaustive, full proactive verification of the total training steps to capture stealthy, parsimonious backdoors, our approach drastically minimizes this burden. By modeling our oracle after PoTS—which typically incurs a 30\% overhead per step—we demonstrate that invoking LV at just a 10\% frequency introduces merely 2.76\% overall computational overhead (192.54 s total vs. 187.37 s base). This confirms the practical efficiency of our defense.


\section{Discussion \& Future Work} \label{sec:discussion}

We formalize the inherent resilience of neural networks against persistent backdoors via continuous benign updates during community training. This resistance is fundamentally driven by catastrophic forgetting \coloredcitep{kemker2018measuring}, wherein the steady influx of benign updates shifts the training distribution and progressively overwrites injected malicious features. By repurposing this natural backdoor absorption as an active structural defense, we demonstrate how sequential honest updates can reliably eradicate adversarial triggers prior to deployment \coloredcitep{du2019lifelong}. 

\noindent \textbf{Alignment with Modern AI Training.} Recent evidence confirms that large language models are highly vulnerable to parsimonious poisoning attacks requiring minimal samples \coloredcitep{souly2025poisoning, hubinger2024sleeper}. Our findings corroborate this, proving that an adversary’s optimal strategy is to minimize their injection lifespan ($\tau$) relative to the network's absorption lifespan ($\pi$) \coloredcitep{zeng2023narcissus}. This asymmetric dynamic renders static auditing computationally unscalable, necessitating the dynamic, absorption-aware defenses. Complementary research demonstrates that distributed training naturally dilutes localized malicious updates \coloredcitep{murturi2023community, hegedHus2021decentralized}. Because this inherent attenuation aligns perfectly with emerging decentralized paradigms—like DeepMind’s decoupled DiLoCo architecture \coloredcitep{douillard2026decoupled}—our framework may therefore serve as a principled basis for extending backdoor defenses to such settings, where centralised oversight is architecturally precluded.

\noindent \textbf{Future Directions.} Because our formulation assumes a central owner orchestrating the verification oracle, future iterations must decentralize this process via consensus protocols to eliminate the single point of failure \coloredcitep{shayan2020biscotti, abbaszadeh2024zero}. Moreover, since attack viability hinges on rapid injection outpacing slow absorption, future defenses should investigate active-learning strategies that artificially accelerate catastrophic forgetting for anomalous updates, actively minimizing the backdoor's absorption lifespan $\pi$ \coloredcitep{du2019lifelong}.

\section{Conclusion} \label{sec:conclusion}

In this work, we transform the natural backdoor absorption inherent to sequential AI training into a rigorous defense mechanism for community-trained models. By formalizing the stochastic competition between adversarial injection and benign absorption as a time-inhomogeneous Markov chain, we derive practical bounds on attack success probabilities and demonstrate that augmenting natural absorption with a strategic defense forces this probability to asymptotically collapse to zero. Empirical evaluations validate this, showing that pairing natural decay with a minimal 10\% lazy verification budget definitively eradicates stealthy backdoors without degrading benign utility. Ultimately, our framework provides mathematically sound and actionable guidelines, enabling model owners to deploy computationally efficient and provably secure auditing strategies across decentralized, safety-critical AI systems.


\bibliography{references}

\end{document}